\documentclass[a4paper,fleqn]{cas-dc}

\usepackage[authoryear]{natbib}
\def\tsc#1{\csdef{#1}{\textsc{\lowercase{#1}}\xspace}}
\tsc{WGM}
\tsc{QE}
\tsc{EP}
\tsc{PMS}
\tsc{BEC}
\tsc{DE}
\begin{document}
\let\WriteBookmarks\relax
\def\floatpagepagefraction{1}
\def\textpagefraction{.001}

\shorttitle{Surface Current and Shear Retrieval from HF Radar}

\shortauthors{G Hasson et~al.}

\title [mode = title]{Retrieval of Ocean Surface Currents and Vertical Shear from HF Radar Observations}
%\tnotemark[1,2]

% \tnotetext[1]{This document is the results of the research
%    project funded by the National Science Foundation.}

% \tnotetext[2]{The second title footnote which is a longer text matter
%    to fill through the whole text width and overflow into
%    another line in the footnotes area of the first page.}

\author[1]{G Hasson}[
%type=editor,
%                        auid=000,bioid=1,
%                        role=Researcher,
%                        orcid=0000-0001-7511-2910
                        ]

\cormark[4]


%\ead{giorahasson@mail.tau.ac.il}

%\ead[url]{www.cvr.cc, cvr@sayahna.org}

\credit{Methodology, Software, Data Curation, Writing}

\affiliation[1]{organization={School of Mechanical Engineering, Tel-Aviv University},
    %addressline={Radarweg 29},
    city={Tel-Aviv},
    % citysep={}, % Uncomment if no comma needed between city and postcode
    postcode={6997801},
    country={Israel}}

\affiliation[2]{organization={Université Paris-Saclay, Ecole Normale Supérieure Paris-Saclay},
    city={Gif-sur-Yvette},
    % citysep={}, % Uncomment if no comma needed between city and postcode
    postcode={91190},
    country={France}}

\author[2]{S St{\o}le-Hentschel}[%style=chinese,
       orcid=0000-0002-8081-9782
]
\ead{Susanne.Stole-Hentschel@proton.me}
\credit{Conceptualization, Methodology, Supervision, Funding acquisition, Writing}

% third author
\author[3]{T Vre\'{c}ica}[%style=chinese
]
%\ead{vrecica@gmail.com}
\credit{Non-linear theory, Writing, Funding acquisition, Writing – review and editing}
% fourth author
\author[4]{V Shrira}[%
   % role=Co-ordinator,
   % suffix=Jr,
   ]
%\ead{\href{mailto:v.i.shrira@keele.ac.uk}\textbf{v.i.shrira@keele.ac.uk}}
%\ead[URL]{www.sayahna.org}

\credit{Supervision, Writing, Methodology}

\affiliation[3]{organization={Mathematical Institute of the Serbian Academy of Sciences},
addressline={Knez Mihailova 36, P.O. Box 367},
city={Belgrade},
    postcode={11000},
    country={Serbia}}

\author%
[1]
{Y Toledo.}
\cormark[1]
%\ead{toledo@tauex.tau.ac.il}
%\ead[URL]{https://web2.eng.tau.ac.il/wtest/toledo/wp/}

\credit{Conceptualization of the study, Methodology, Supervision, Funding acquisition, Radar Resources}

\affiliation[4]{organization={School of Mathematics and Computing, Keele University},
city={Keele ST5 5BG},
country={UK}}

\cortext[cor1]{Corresponding author}
%\cortext[cor2]{Principal corresponding author}

%\fntext[fn1]{This is the first author %footnote. but is common to third
 % author as well.}
%\fntext[fn2]{Another author footnote, this %is a very long footnote and
  %it should be a really long footnote. But this footnote is not yet
  %sufficiently long enough to make two lines of footnote text.}

%\nonumnote{This note has no numbers. In this work we demonstrate $a_b$
  %the formation Y\_1 of a new type of polariton on the interface
  %between a cuprous oxide slab and a polystyrene micro-sphere placed
  %on the slab.
  %}

\begin{abstract}
The upper few meters of the ocean play a key role in air–sea exchanges of momentum and energy. Two important properties of this layer are the vertical shear of current velocity and the surface velocity. Vertical shear reflects momentum transfer, while the surface current velocity is essential for modeling transport processes such as pollutant dispersion. Direct measurements are challenging, and methods that provide continuous, real-time, and spatially extensive observations are highly desirable.

This proof-of-concept study presents a new approach for estimating the radial component of surface velocity and the vertical shear of sea surface currents using a single-frequency high-frequency (HF) radar. The method is based on estimating the two first-order Bragg peaks in the backscattered radar spectrum which exhibit asymmetric frequency shifts in the presence of shear. The integral shear is retrieved from the frequency difference between the two peaks, while a new spectral processing technique enables correction of the surface velocity estimate for the effect of shear. The approach was tested using data from a WERA HF radar deployed off Ashkelon (Israel), and validated against simultaneous acoustic Doppler current profiler (ADCP) measurements at lower depths. The retrieved shear magnitudes and surface velocities show consistent trends with ADCP and wind observations, indicating that the method is applicable under strong-wind conditions for HF radar systems providing high-quality Doppler spectra.
\end{abstract}

% \includegraphics{figs/grabs.pdf}

\begin{highlights}
\item A proof-of-concept for extracting the vertical shear and current magnitudes at the sea surface using the first-order Bragg peaks of a single-frequency HF-radar
\item Field campaign shows the methodology's applicability under strong wind conditions
\item A preliminary study of nonlinear wave effects indicates significant Doppler shifts for strong wave conditions
\end{highlights}

% Keywords

\maketitle

\section{Introduction}
\noindent Processes in the upper few meters of the ocean play a key role in controlling the exchange between the ocean and the atmosphere \cite{donelan1990air}. The vertical profile of this upper layer current  encapsulates information on the turbulence intensity distribution, the latter  is of great importance to the exchange of heat and momentum fluxes at the air-water interface. This significance is reflected in recent introduction of  two `\emph{Essential Ocean Variables': Surface Current and Ocean Surface Stress}\footnote{https://goosocean.org/what-we-do/framework/essential-ocean-variables/}.

HF radars are a unique remote sensing tool that can continuously provide near real-time surface current maps across large areas for various purposes \cite[see e.g.,][]{Lorente2022, Reyes2022}. In this proof-of-concept work, we lay the foundation of a novel way of remote sensing of the subsurface shear flow and its velocity at the surface using a single HF radar by utilizing the frequency shifts in the first-order peaks in the HF radar echo spectrum. These peaks, which result from radar signal reflection from the sea surface, were first observed by \cite{crombie1955doppler}. In the Bragg scattering regime, relevant for the HF and VHF radars, the peaks are due to resonant scattering by the waves of half the wavelength of the emitted electromagnetic wave \cite{lipa1986extraction}. The discrepancy between the observed frequency of the Bragg lines and the frequency given by the linear dispersion relation for the resonant surface gravity wave is  the Doppler shift due to the current. In terms of velocities this  discrepancy is usually interpreted as velocity of the surface current, although this would have been true only in the absence of vertical shear.

Here, we propose an extension of the use of HF radars, which at present are primarily used to provide surface current maps over large areas  neglecting the shear, e.g. \cite{gurgel1999wellen}. More advanced approaches that allow also the extraction of shear current from HF radar measurements have been developed. As outlined in \cite{stewart1974hf}, the utilisation of multiple HF frequencies simultaneously, with each frequency observing distinct Bragg waves, enables the vertical probing of the shear current. This method for describing waves in the presence of a vertically shearing current is based on approximations based on the Rayleigh equation  \cite[see, e.g.][for a more improved approximation and an extension to a simplified wave-action equation]{shrira1993surface, Quinn2017} Although this method is relatively simple to implement, it requires having two or more operating frequencies, which is not common in HF radars.  \cite{shrira2001remote,ivonin2004validation} extended the usage of HF radars by utilizing the second-order and corner reflection peaks in the echo  Doppler frequency spectra. These additional peaks result from surface waves of different wavelengths, and thus provide a way of probing  the  shear while employing a single frequency radar. However, the secondary peaks are much weaker than the first order ones, which sometimes makes them difficult to discern.  
 
Here we propose a novel approach for determining vertical shear  and surface  current utilizing a single-frequency radar and only the  first-order peaks in the  radar echo  Doppler frequency spectra. To this end we exploit the observation that in the presence of shear, the positions of two first-order peaks resulting from Bragg waves propagating in the opposite directions (toward and away from the radar)  are not symmetric with respect to the radar frequency. By inferring the associated phase velocities, we demonstrate that the vertical shear and current velocity at the surface can be estimated. Although the differences between the two phase velocities are relatively minor, we show that it is nevertheless feasible to estimate the shear under strong wind conditions. The results are consistent with the current profile somewhat lower in the water column, as measured simultaneously by an Acoustic Doppler Current Profiler (ADCP). Thus, this work establishes the foundation for large-scale shear estimation using a single HF radar.

 Although the proposed approach is based upon dispersion properties of linear surface waves on shear flow, we also examine the significance of nonlinear effects. Nonlinear corrections to dispersion relation \cite{zakharov1999statistical} are of second order in wave steepness, and, therefore, are often neglected. As previously stated, the difference in velocities of the two peaks is small, and in view of this discarding these effects is not necessarily justified. A preliminary analysis indicates that the corrections to dispersion manifest in modulations of surface current (under the assumption of linear current profile).  It is demonstrated that for energetic wave fields nonlinear corrections can be of comparable magnitude to the linear contribution, but can be safely neglected for relatively calm seas.

\begin{figure}[h]
    \centering
    \centerline{\includegraphics[width=\linewidth]{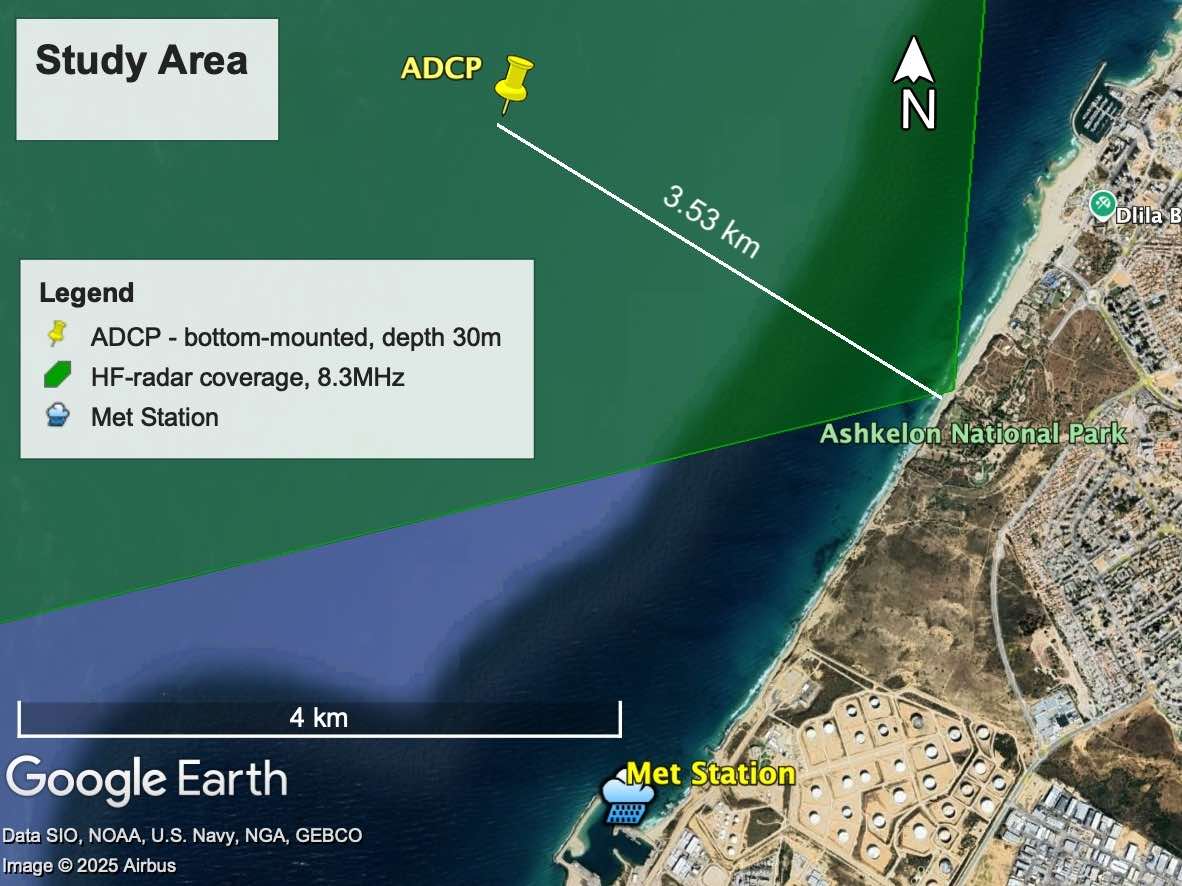}}
    \caption{Satellite image from Google Maps, overlaid with key features relevant to the study area. The HF station is marked in red. The shaded green area represents the coverage area of the HF station in the nearshore region (with maximal range of $\sim$180km in ideal conditions). Additionally, the location of the ADCP (bottom depth of 30m) can be seen in front of the radar station, providing information on the current profile and the waves. Lastly, the location of the meteorological station is indicated to the south of the HF station, providing data on local weather conditions.}
    \label{fig:region}
\end{figure}

The paper is organized as follows. The HF radar and ADCP measurements are described in section 2, together with a detailed description of the data processing. The novel methodology of estimating the shear and current velocity at the surface  is presented in section 3. In section 4, preliminary results on the effects of non-linearity are outlined. The results are presented in section 5, showing how the upper-layer shear from our new estimate responds to the changes in the environmental conditions. With a certain degree of latency, the fluctuations in upper-layer shear are also discernible in the ADCP data a few meters below the sea surface. The final section of the work, section 6, presents the conclusions and a brief discussion.

\section{Methodology} \label{sec:materials}
\noindent  The study utilized a WERA HF radar situated within the Ashkelon National Park at the coast of the  Mediterranean Sea, however the methodology of the proposed approach is not radar specific.  The station consists of $4$ transmitting antennae operating at the frequency of $f_0=8.3$ MHz in a square shape and an array of $12$ receive antennas with a distance of $9$ m ($\lambda_{\mathrm{EM}}/2$) between adjacent antennae. The system is configured to take a measurement every $20$ minutes, with the data averaged over $17.75$ minutes (with an additional $2.25$ minutes for premeasurement calibration). The system generates both current maps and raw spectra, which are then transmitted to a local server at the station. We use the raw spectra to infer the shear and current velocity at the surface. The analyzed field data was acquired from 23 March to 7 April 2021, when three storms hit the area. The method will first be presented with a focus on the longest storm that occurred March 23 to March 25. As conventional method to estimate shear is not available, we validate the presented measurements with ADCP and wind measurements. The ADCP was only avaialbe during the chosen period which are all presented here. The wind measurements stem from a  nearby meteorological station (Figure \ref{fig:region} positioned at 31.6692° N; 34.5021° E, depth 30 m distance to shore 3.53 km), providing wind speed, direction and gustiness. With the given depth, the measurements were taken in deep water regime, however, the method is expected to be applicable in intermediate depth and possible shallow water if the distance to the radar antenna ensures high quality measurements (starting from $~2.5$ km in our case).

To extract separate information on the two first-order peaks, we employed the WERA FORTRAN package to generate backscatter spectra in the original polar frame of reference ($r$, $\theta$) from the backscattered signal of each receiving antenna. Example spectra a provided in the supplementary material. To minimize back-scatter noise, a location at a close range and near the HF antennae bore-sight was chosen\footnote{At distance $3.53$ km and angle range of ($-4^\circ$, $5^\circ$) contains area of about $1 \mathrm{km}^2$.} (see Figure \ref{fig:region}). The finding of the peak location, $\bar{f}$, is analogous to the approach currently employed by WERA as standard. For a spectrum containing NFFT data points, a window of size $K=$ NFFT$/64$ is extracted around the detected first order peaks. A weighted average of each element within the window ($f_n$) is then calculated using the power ($P_n$) as the weighting function:
\begin{equation} \label{eq:freq_location}
    \bar f =  \frac{\sum_{n=1}^{K}f_n P_n}{\sum_{n=1}^{K}P_n} \quad .
\end{equation}

In contrast to the common approach, we estimate separately the frequency peaks resulting from offshore propagating waves ($f_{-}>0$) and onshore propagating waves ($f_{+}<0$). The respective Doppler shifts ($c_{-}$ and $c_{+}$) are calculated from the measured frequency shift with respect to the radar frequency $f_0$ using a two-way relativistic approach  (see full derivation in the Appendix):
\begin{subequations} \label{eq:dop_freq_to_vel}
\begin{align}
    c_{\mathrm{-}} =  \frac{C_e\left( f_0 - f_{\mathrm{offshore}}\right)}{f_0 + f_{\mathrm{offshore}}} \quad ,\\
    c_{\mathrm{+}} =  \frac{C_e\left( f_0 + f_{\mathrm{onshore}}\right)}{f_0 - f_{\mathrm{onshore}}} \quad ,
\end{align}
\end{subequations}
where $C_e$ is the speed of light. The continuous  HF radar measurements allow  a time series extraction of $c_{-}$ and $c_{+}$.

To estimate the shear from the difference between the offshore and onshore propagating waves celerity measurements, it is necessary first to ensure that the error \cite{essen2000accuracy} is smaller than the discrepancy between the measurements. To determine the confidence level, we utilize the maximum error $  \Delta$ and the specified resolution, based on the WERA's accuracy estimation:
\begin{equation} \label{eq:confidence}
    \Delta =  \max(dv, \sqrt{\sigma^2}) \quad , \quad
    \end{equation}
where $\sigma$ is the standard deviation and  $dv$ is the measurement resolution derived from the radar transmission configuration,
\begin{equation} \label{radar_resolution}
    dv = \frac{1}{2 \mathrm{N} T_{\mathrm{chirp}}}\quad .
\end{equation}
The term $N T_{\mathrm{chirp}}$ denotes the duration of the radar measurement, the longer it is, the finer the resolution. The value of $N T_{\mathrm{chirp}}=17.75$ minutes results in a frequency resolution of 0.001 Hz, corresponding to a velocity of approximately $1.7$ cm/s. The accuracy of the measurement is defined by the standard deviation $\sigma$, 
\begin{equation} \label{eq:WERA_variance}
    \sigma=\sqrt{\frac{\sum_{n=1}^K f^2(n) \mathrm{SNR}(n)}{ \sum_{n=1}^{K} \mathrm{SNR}(n)} - \bar f^2\quad }.
\end{equation}
Here, $\mathrm{SNR}$ is the signal to noise ratio obtained from the measurement. The standard deviation is calculated using the same window size as the peak detection estimate.

 The vertical profile of the current in intermediate and low layers was measured using the NORTEK Signature 500, an Acoustic Doppler Current Profiler (ADCP) deployed at a distance of 3.5 km from the radar station and at an angle of 1 degree from the bore sight of the antennae array (Figure \ref{fig:region}). The deployment included external battery canisters that allowed logging 5-beam bursts continuously. This enabled better accuracy in fitting the time-series comparisons between the two instruments. It measured the north-south and east-west components of the horizontal velocity in the depth range $30$ to $5.7$ m with a step of 0.2 m (Figure \ref{fig:solution_fig} right panel). The accuracy of the ADCP measurements depends on the wave height. The total error of the ADCP measurements was estimated as the sum of the standard deviations of the measurements during the time interval. To account for the sampling variability of ADCP and HF radar, the measurements were averaged over a one hour period.  For the comparison to the shear current obtained from the radar, the ADCP measurements were projected onto the look direction of the radar.

%The analyzed field data was acquired from 23 March to 7 April 2021, when a storm hit the area. In addition to the HF radar and ADCP measurements, the available data includes measurements from a  nearby meteorological station (Figure \ref{fig:region}), providing wind speed, direction and gustiness.

\section{Theoretical estimates}
\noindent The utilized WERA radar transmits a \emph{Frequency Modulated Continuous Wave} (FMCW) with the radar  frequency $f_0$ \citep{gurgel1999wellen}. Due to the dominance of the Bragg resonance mechanism, a strong signal is back-scattered from the surface waves at Bragg wavenumber $k_B$. These surface waves propagate to and from the radar, with the respective phase velocities $c_+$ and $c_-$. The relation between the electromagnetic waves and the Bragg waves is defined by their respective wavelengths, $k_B=2K_0=4\pi f_0/C_e$. HF radar signal can be in resonance  with water waves of wavelengths in the range $10-100$m which are measured sufficiently far from the shore, so that the deep water dispersion relation can be assumed:
\begin{equation} \label{dispersion_relation}
    \omega_0=c k,
\end{equation}
where $c$ is the water wave celerity, $k$ is the water wavenumber and $\omega_0$  in is the angular frequency of the water wave. The frequency shift  $f_B=\omega/2\pi$ that is induced by the current is detectable in the frequency spectrum. It is calculated as deviation between measured frequency peaks and transmitted frequency: %The radar measures the frequency shift $f_B=\omega/2\pi$ corresponding to the transmitted frequency. This means that, for a given radar configuration, you can predict exactly where the Doppler peaks will appear in the radar spectrum.:
\begin{equation} \label{dispersion_relation_with_current}
    \omega=\omega_0 + \textbf{k} \cdot \textbf{U} \quad ,
\end{equation}
where \textbf{U} is the current vector and \textbf{k} is the wave-number vector. Equation \ref{dispersion_relation_with_current} suggests that both peaks experience the same shift. This approach was further developed by \cite{barrick1972first} and is now a standard method for extracting current maps.

However, the assumption of a depth--uniform current is difficult to justify in the presence of strong vertical shear. Shear has different effects on waves propagating to and from the radar: a wave propagating in the direction of a shear current has a different phase velocity than the wave propagating in the opposite direction. To evaluate the effect of the shear on the waves, the Rayleigh equation with a free surface boundary condition can be used to analytically or numerically extract a relationship between the current vertical profile  $U(z)$, and celerity $c$:
\begin{subequations} \label{eq:Rayleigh_equations}
\begin{align}
  w'' + \left[\frac{U''}{c-U} - k^2 \right]w  = 0,  \label{eq:Rayleigh1} \quad  \\
    w' =\left[\frac{g}{(U-c)^2}+\frac{U'}{U-c} \right]w  \quad \textrm{at} \quad z=0, \quad   \label{eq:Rayleigh2} \\
    w=0   \quad \textrm{at} \quad z\rightarrow-\infty. \quad
\end{align}
\end{subequations}
Here, $w$ is the vertical component of the $3D$ velocity vector and prime denotes derivation with respect to vertical coordinate $z$, $g$ is gravitational acceleration.  The simplest model of shear current is the model assuming constant shear:
\begin{equation} \label{eq:constant_shear}
    U(z)=\alpha  z + \beta \quad 
\end{equation}
where $\alpha$ denotes the shear parameter and $\beta$ the surface current. The current affecting the Bragg wave relates to the relatively thin uppermost water layer as to $O(1)$ it the current's influence is weighted by $e^{2kz}$ \citep[see][]{stewart1974hf}. Hence, $\alpha$ models the near-surface shear.
%For simplicity, this case will be studied in this letter, although other approximations such as \cite{shrira1993surface}, \cite{ellingsen2017approximate} can be called upon within the framework of the approach.

On solving the Rayleigh equation Eq.~\ref{eq:Rayleigh_equations} for the adopted constant shear parametrization Eq.~\ref{eq:constant_shear} we find the dependence of $c_+$ and $c_-$ on the current parameters $\alpha$ and $\beta$:
\begin{subequations} \label{eq:Ray_lin_sol}
\begin{align}
    c_+ = \frac{-\alpha+\sqrt{4 g k + \alpha^2}-2 k\beta}{2k}\quad , \\
    c_- = -\frac{\alpha+\sqrt{4 g k + \alpha^2}+2 k\beta}{2k} \quad ,
\end{align}
\end{subequations}
where $k$ is the resonant Bragg wavenumber specified by the radar transmission frequency $f_0$, and $c_+$ and $c_-$ are extracted from the radar measurement. It is straightforward to explicitly express $\alpha$ and $\beta$, the parameters of the model,  in terms of the directly measurable characteristics:
\begin{subequations} \label{eq:constant_lin_sol}
\begin{align}
    \alpha = \pm \sqrt{-4 g + (c_+-c_-)^2k^2} \quad ,\label{eq:Alpha}\\
    \beta = \frac{1}{2} \left(-c_+-c_- \pm \sqrt{\frac{-4 g + (c_+-c_-)^2 k}{k}}\right) \quad .\label{eq:Beta}
\end{align}
\end{subequations}
The ambiguity of the solutions for $\alpha$ and $\beta$  can be resolved using the direction of the current. For the WERA system the convention imposes that a positive deviation of the peaks ($c_+-c_->0$) leads to a peak shifted to the right of the Bragg frequency ($f_B$) and the corresponding current is directed away from the radar and  $\alpha,\beta$ are positive. Under the assumption that the shear is weak, the above equations can be simplified. Formulations for $c_{+}$ and $c_{-}$ accurate up to $O(1)$ in shear are defined as:
\begin{equation}
    c_{\pm} \approx \frac{- \alpha}{2k} \pm \sqrt{g/k}(1+\alpha^{2}/8gk) - \beta.
\end{equation}
These equations show that, in the linear profile approximation, there is no contribution of shear at $O(\alpha)$ to the difference of phase velocity $(c_{+}-c_{-})$. This simplified form clearly demonstrates at which orders different effects come into play, however equation 11 is used throughout the paper.

\section{Revisiting nonlinearity}
\begin{figure*}[ht]
\centering
    \includegraphics[width=0.99\linewidth]{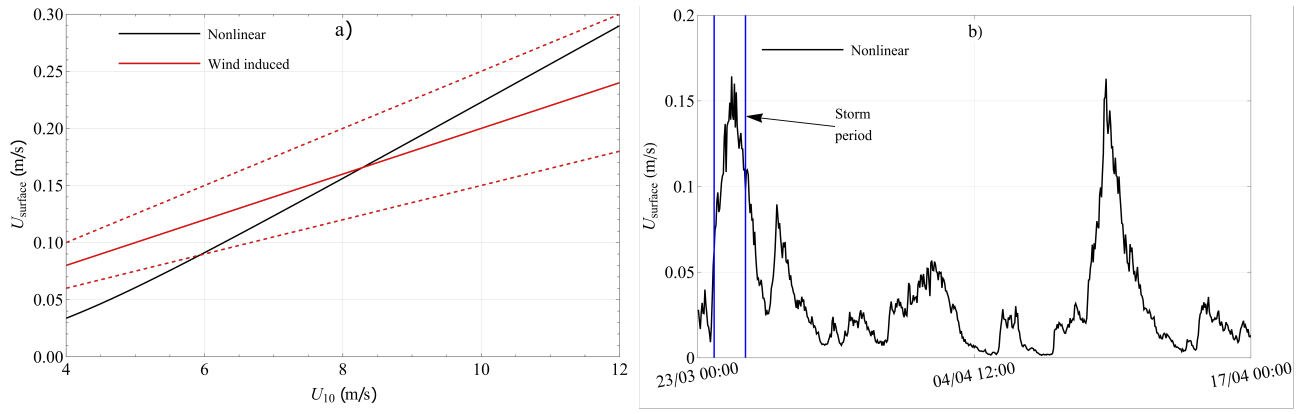}
  \caption{Nonlinear effects on the surface current ($\beta$). a) A comparison of the surface currents induced by wind and by taking into account nonlinearity in the nonlinear dispersion relation computed for a Pierson-Moskowitz spectrum. Results are shown as a function of $U_{10}$) (wind speed at 10m height).  b) Nonlinear correction to surface current computed for waves whose wavenumbers corresponding to radar wavelengths. The correction is computed using wave spectra measured by an ADCP, from third of March to seventeenth of April. }
\label{fig:NL}
\end{figure*}
In this section, we consider the effects of the nonlinear contribution to the surface wave dispersion relation on the vertically sheared currents. We only consider resonant four-wave nonlinear interactions, as discussed in \cite{zakharov1999statistical}. It is further assumed that the effects of the shear on nonlinearity are small, and to leading order a potential nonlinear theory can be used. A full derivation and analysis of the combined effects of nonlinearity and shear goes beyond the scope of this work.

%Concurrent ADCP in-situ measurements of wave spectra are used in this paper to estimate the nonlinear dispersion relation.
Waves travel faster in the mean direction of the wave propagation, while waves travelling in the opposite direction of the main wave system are slower \cite{masuda2000nonlinear}. The Stokes wave solution valid for a single monochromatic  wave is a classic example of this phenomenon. In case of an arbitrary wave spectrum  the wavenumber for which the nonlinear dispersion is calculated is determined by the radar frequency, and for broad-banded wave systems the corresponding nonlinear angular frequency ($\Omega$) is
\begin{equation}
    \Omega (\mathbf{k})=\omega(\mathbf{k}) + 4 \pi^{2} g \int T(\mathbf{k},\mathbf{k_{1}}) \frac{S(\mathbf{k_{1}})}{\omega(k)} \mathrm{d}k_{1},\label{eq:NL_dispersion}
\end{equation}
where $S(\mathbf{k})$ is the energy wave spectrum, and the interaction kernel $ T(\mathbf{k},\mathbf{k_{1}})$ t is a lengthy function defined in terms of wavenumbers and frequencies \cite{zakharov1999statistical}, although it is dramatically simplified for the unidirectional case \cite{lavrova1983transversal}:
\begin{eqnarray}
    T(k,k_{1})=\frac{1}{2 \pi^{2}} \begin{cases}
      k^{2}k_{1} & k<k_{1},\\
      k k_{1}^{2} & k>k_{1}.
    \end{cases}
\end{eqnarray}
This enables us to express wave velocity with account of nonlinear effect, $c_{nl}$,  in terms of the linear one $c_{lin}$
\begin{equation}
    c_{nl}=c_{lin}+2 \left(\int_{0}^{k} k_{1} \omega_{k_{1}} S(k_{1}) \mathrm{d}k_{1} + \int_{k}^{\infty} k \omega_{k_{1}} S(k_{1}) \mathrm{d}k_{1}\right).
\end{equation}
Thus, the specific wavenumber is selected by the Bragg resonance, but the corresponding frequency we encounter in the Doppler echo spectra  differs somewhat  from the linear frequency, which is essential for interpreting the observations. 

When these corrections  to phase velocities are applied under the square roots in equation \ref{eq:constant_lin_sol} they effectively cancel each other out. The absolute contribution to $c_{+}$ and $c_{-}$ are nearly identical and thus $(c_{+}-c_{-})_{nl} \approx (c_{+}-c_{-})_{lin}$. However, there is a non-negligible correction to the $\beta$ term in equation \ref{eq:constant_lin_sol}b. In the linear profile model  this corresponds to the surface current.

For brevity we will refer to the nonlinear correction to frequency as the Stokes-type correction. A comparison of magnitudes of the  Stokes-type correction (expressed through $-c_{1} - c_{2}$) and wind induced surface currents ($\beta$) is given in Figure \ref{fig:NL}a.  The nonlinear correction computed for the Pierson-Moskowitz spectra, which is  a function of wind speed $U_{10}$ (wind speed at 10m above the mean water level):
\begin{equation}
    S(k)=\frac{0.004}{k^{3}}exp \left(\frac{-0.554 g^{2}}{U_{10}^{4} k^{2}} \right).
\end{equation}
An analytic solution for the nonlinear dispersion relation for such a spectrum was derived in \cite{stuhlmeier2019nonlinear}. Measurements of wind-induced surface current were performed in \cite{kudryavtsev2008vertical}, and as a crude approximation we assume here that the wind-induced surface current is $\approx$ 2\% of $U_{10}$. A comparison of these two effects is given in Figure \ref{fig:NL}a as a function of $U_{10}$. It is found that, overall,  the magnitudes are comparable, at least for developed seas.

The nonlinear contributions to $\beta$ were computed using wave spectra obtained via in-situ ADCP measurements under potential flow assumptions \cite[see][for effects of the shear on spectrum extraction which are not required in this case]{Soffer_ADCPshear_2023}. It is assumed that the wave field propagates in the mean wind direction. The surface currents are also computed, under the assumption of no shear (measured departure from linear solution). The magnitudes of the nonlinear correction and of the measured current are shown in Figure \ref{fig:NL}b. The rapid oscillations of the measured velocities occur during periods with little wave energy, during which the radar accuracy is significantly lower. It should be kept in mind that the assumption of unidirectional spectra amplifies the effect of nonlinearity, and the assumption of no shear reduces the magnitude of the measured surface current. Nonetheless, these results show that the contribution of nonlinearity, i.e. the  Stokes-type correction, is indeed significant for estimation of surface currents (e.g. the storm period of 23-25th of March).

\begin{figure}[h]
    \centering
    \includegraphics[width=1.1\linewidth]{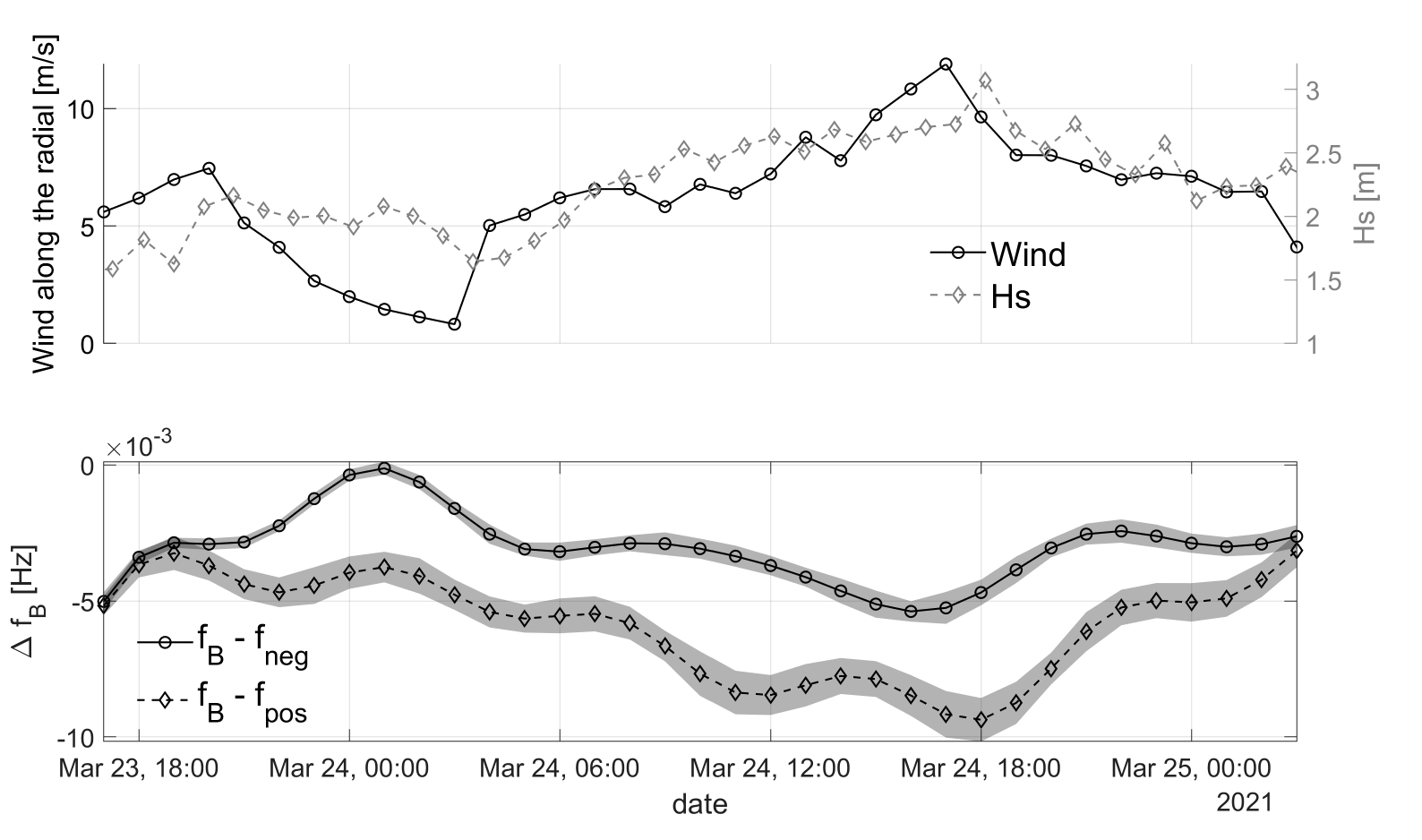}
    \caption{Measurements over the presented period. Upper panel: Significant wave height (dashed gray) and absolute value of the wind projected onto the HF direction. The wind is blowing towards the radar station (the shore) during the storm. Lower panel: Deviation of the extracted peak location from the Bragg frequency for the positive Doppler shift (dashed) and negative Doppler peak (solid). Under the constant current assumption the deviation should be the same.  The deviation of the peak location is greater than the estimated error which is shown in grey.}
    \label{fig:peaks_deviation}
\end{figure}

\section{Results} \label{result_section}

% {Figures} may be included using the command,\linebreak
% \verb+\includegraphics+ in
% combination with or without its several options to further control
% graphic. \verb+\includegraphics+ is provided by {graphic[s,x].sty}
% which is part of any standard \LaTeX{} distribution.
% {graphicx.sty} is loaded by default. \LaTeX{} accepts figures in
% the postscript format while pdf\LaTeX{} accepts {*.pdf},
% {*.mps} (metapost), {*.jpg} and {*.png} formats.
% pdf\LaTeX{} does not accept graphic files in the postscript format.

% \begin{figure}
% 	\centering
% 		\includegraphics[scale=.75]{Fig1.pdf}
% 	\caption{The evanescent light - $1S$ quadrupole coupling
% 	($g_{1,l}$) scaled to the bulk exciton-photon coupling
% 	($g_{1,2}$). The size parameter $kr_{0}$ is denoted as $x$ and
% 	the \PMS is placed directly on the cuprous oxide sample ($\delta
% 	r=0$, See also Table \protect\ref{tbl1}).}
% 	\label{FIG:1}
% \end{figure}

% The \verb+table+ environment is handy for marking up tabular
% material. If users want to use {multirow.sty},
% {array.sty}, etc., to fine control/enhance the tables, they
% are welcome to load any package of their choice and
% {cas-dc.cls} will work in combination with all loaded
% packages.

% \begin{table}[width=.9\linewidth,cols=4,pos=h]
% \caption{This is a test caption. This is a test caption. This is a test
% caption. This is a test caption.}\label{tbl1}
% \begin{tabular*}{\tblwidth}{@{} LLLL@{} }
% \toprule
% Col 1 & Col 2 & Col 3 & Col4\\
% \midrule
% 12345 & 12345 & 123 & 12345 \\
% 12345 & 12345 & 123 & 12345 \\

% \bottomrule
% \end{tabular*}
% \end{table}

\noindent The proposed method was applied to the data from March 23rd to 25th and provides hourly estimates of the frequency shifts and the associated shear. During the selected period, the wind speed was above 6~m/s for most of the time and the significant wave height in the range of 1.5 -- 3~m (see upper panel in Figure \ref{fig:peaks_deviation}). The frequency shift between the extracted peak frequency and the unperturbed Bragg frequency is shown in the lower panel of Figure \ref{fig:peaks_deviation}.  The estimated error (calculated according to equation \ref{eq:WERA_variance}) is noticeably  smaller than the difference between the observed frequency shifts. Furthermore, we stress that the frequency shifts for waves propagating towards and away from the radar follow a distinct pattern. These observations show that a shear is present in the current, since assuming a constant current  would result in identical frequency shifts.

\begin{figure*}[t]
    \centering
    \includegraphics[width=1\textwidth]{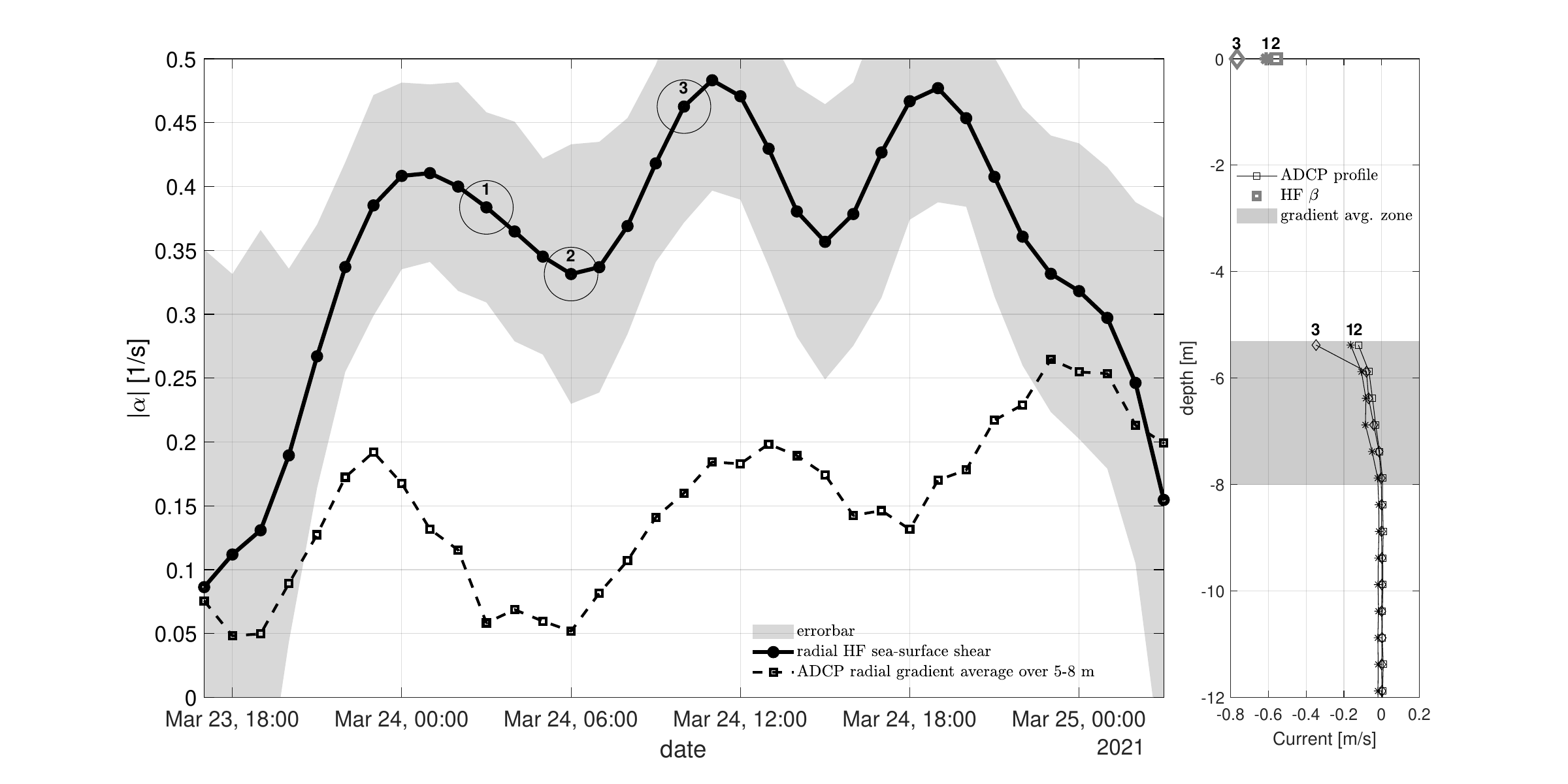}
    \caption{Results obtained from the HF and an ADCP for the depicted period. Left panel: comparison between the HF sea-surface shear (our method) to the average gradient of the current profile extracted by an ADCP over the depth $5$ to $8$ meters together with the surface current value from the radar, both estimations share the same trends. The trend of the shear penetration is shown to have a reasonable physical agreement with a reduced shear and delay in penetration to deeper region, with some discrepancy in the first lag possibly due to large errorsbars in the beginning of the measurement section.  Right panel: three representative radial current profiles extracted by ADCP, the time of each profile is circles on the left panel. Gray shaded area present the averaging zone to create the shear estimation from the ADCP profile.}
    \label{fig:solution_fig}
\end{figure*}

The shear and the sea surface current are then estimated by plugging  the measured difference between the frequency shifts into equations \ref{eq:Alpha} and \ref{eq:Beta}. The calculated shear and the corresponding confidence bounds are shown in the left panel of figure \ref{fig:solution_fig}. No direct measurements of the current profile in the uppermost ocean layer are available, as such data are notoriously difficult to obtain. For the validation of the HF shear estimates, we are comparing the HF shear estimates with the average gradient of the radial current profile measured  by an ADCP at a somewhat larger depth: The average over depths of $5-8$ meters. The values were chosen to be as close to the surface as possible and still include a couple of measurement points together with the surface current value from the radar. Increasing or decreasing the interval length did not change the results considerably. The comparison revealed two characteristics: First, the shear parameter is generally lower for the ADCP, as it measures at lower depths; the HF, on the other hand, measures higher values, which is indicative of strong shear in the boundary layer of the air-sea interface. Second, the fluctuations of the shear are similar for both  parameters, with temporal shifts for the cusps. The upper layer tends to respond faster to changes in the wind, and it takes some time until the momentum is transferred to lower layers of the water column. An exception to this trend is the first peak in the shear, which occurred around midnight on March 24th. In this instance, the ADCP shear parameter exhibited a peak  before the HF shear parameter.
The details of three time instances from the left panel (see markers 1,2 and 3) were investigated in greater detail in the right panel of Figure \ref{fig:solution_fig}. The surface current ($\beta$) estimated by our method is  indicated at the mean water level. The corresponding radial current profiles over depths of $5-12$ m are shown at their corresponding depths. The behaviour of the ADCP current profiles has a tendency towards the HF surface current. It is also notable that instance number three demonstrates a pronounced shear between the uppermost measurement points of the ADCP and a particularly strong shear and surface current estimated by the HF radar.
\noindent

\begin{figure*}
    \centering
    \includegraphics[width=1\linewidth]{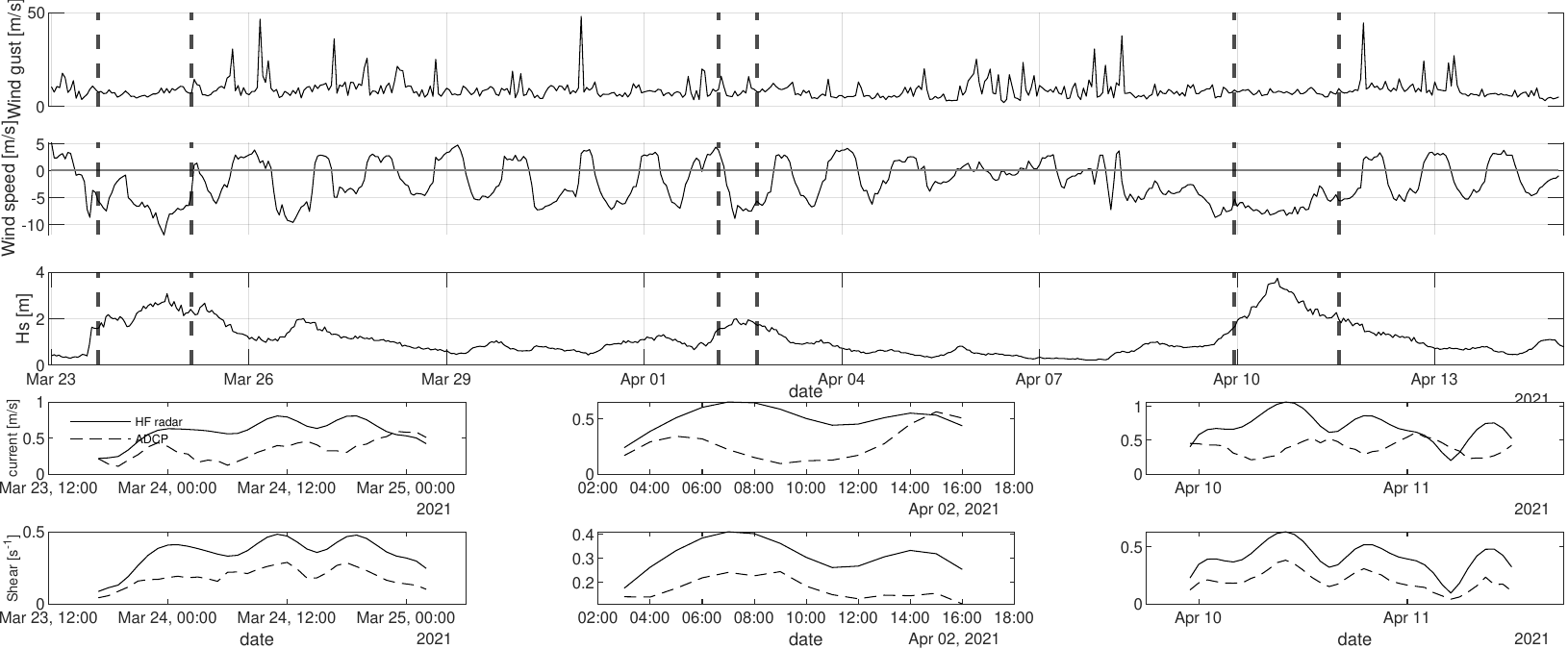}
    \caption{Measurements over the entire period. Vertical dashed lines indicate the intervals where our method succeeds (the first interval is the one for which elaborated results in figure \ref{fig:solution_fig}). Upper full panel: Wind gustiness as measured by a nearby meteorological station, we can see the regions of existing results are characterized by very low wind gustiness. Middle full panel: absolute value of the radial wind speed, all the regions our method was able to succeed, the wind was strong, however, periods of strong wind exist with our method fail to succeed. Lower full panel: significant wave height as measured by an ADCP. Each of the lower couples of sub-panels provide currents (radial surface current $\beta$ and the ADCP's 5m depth current in the same direction) and shear (radial $\alpha$ and the $\beta$ together with ADCP's 5--8m estimated shear in the same direction). We see that our method provides  similar temporal evolution across the storm events for the upper-layer shear $\alpha$, while the expected magnitude difference remains consistent with the different effective depths of the measurements.}
    \label{fig:wind_gust}
\end{figure*}

The presented results were obtained from data of a regional storm with high seas and strong winds (Figure \ref{fig:peaks_deviation} top panel). When applying the method to a longer period, it was noted that the solution of equation~\ref{eq:Alpha} can be complex valued. Figure \ref{fig:wind_gust} shows the wind gustiness, the wind in radial direction and the significant wave height. The vertical dashed lines indicate three intervals for which the proposed method has real valued solutions. As demonstrated in the figure, when the method is applicable, the wind is strong and the gustiness is low and the significant wave height is high. Possible reasons for this are discussed in section \ref{sec:discussion}.

\begin{figure}
    \centering
    \includegraphics[width=0.5\textwidth]{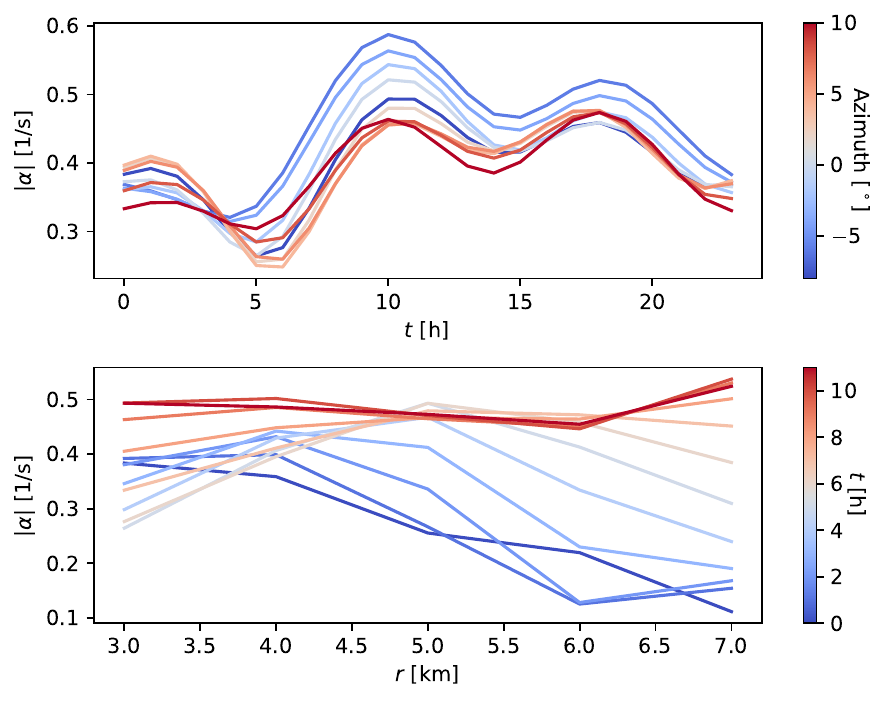}
    \caption{Dependency of shear on azimuth and range. Upper panel: Shear values over time for different azimuth positions; lower panel: Shear values over range at different times.}
    \label{fig:areal_dep}
\end{figure}
Fig.~\ref{fig:areal_dep} presents the measured upper layer radial shear for different azimuth and range over time. The image includes data from March 24$^\textrm{th}$. In the subplot for range only the first 11 hours of the day are shown to avoid overloading the image. The default position for range and azimuth for both images is the one that was used for the comparison with the ADCP (3.3 km at an angle of $0^{\circ}$). Fig.~\ref{fig:areal_dep} shows that shear estimates from adjacent points in space are consistent. The temporal changes in one azimuth position happen simultaneously in other positions while the range plot may be interpreted as a shear wave that propagates in the temporal scale of hours.

\section{Discussion and Conclusion} \label{sec:discussion}
\noindent The presented analysis of a three-week field campaign shows a notable discrepancy between the two first-order frequency peak locations estimated by the HF radar. The difference between the frequency peaks for waves approaching and receding from the radar is indicative of shear in the uppermost layer of the water column. It can be concluded that already existing HF radars can be  used for  probing sea surface shear and current velocity utilizing only the first order peaks. Although the utilized radar was operated in beam forming mode, the method is also expected to work in direction finding mode.  In situ  measurements of the current profile at depths below 5~m  are consistent with the shear and surface current derived from the HF measurements. All considerations were carried out for deep water, but the method is expected to be applicable for finite depth as well.  Our preliminary investigations indicate that spatial measurements in the order of square kilometers are feasible. 

It appears, that the shear can only be retrieved by this method when the wind and wave conditions are appropriate: the approach yields robust solutions when the significant wave height and wind speed in the radial direction are high and gustiness is low. A possible explanation for the influence of gustiness is that it may impede the transmission of the weaker signal of the waves that are propagating away from the radar. Currently, it is not clear to what extent all three parameters are essential for the method. It can be assumed that the significant wave height correlates with a higher signal-to-noise ratio (SNR). However, it is also possible that the significant wave height can be high without measurable shear.  The waves can generate vertical shear via radiation stresses. However, when breaking, they can produce an enhanced mixing and change the air-water momentum fluxes which in-turn reduce the upper-layer shear.   This leaves us with a complex multi-faceted physical behavior in addition to the remote-sensing aspects.

Although the estimated error for the measurements was generally lower than the measured frequency difference between the peaks, it is possible that the adopted error estimate may have overlooked certain contributions that are not significant for the traditional current measurement but have a considerable effect on the shear calculation. Further research is required to enhance our understanding of the methodology limitations and ways to improve it.

Certainly, the validation remains a critical open issue. Further validation will be undertaken in our future work. Due to technical limitations, the depth of the ADCP measurements was restricted to depths five metres from the surface and below. This is a significant constraint for the validation of the uppermost layer. Advanced instruments such as the no-blanking ADCPs and single-point current meter can be employed on moored floating buoys to measure the current profile in the upper 1.5 meters, which includes the actual shear measured by the HF. The principal challenge in utilizing these instruments are their operational and permit requirements. A dedicated expedition should be undertaken in order to obtain accurate measurements of strong shear. Another option might be a route of comparison to the complementary and already validated approach of \cite{shrira2001remote} which was based on utilizing second order peaks.

At the most fundamental level, we have identified  the displacement of the first order peaks due to shear effects. We conclude, that the proposed method can be used for estimating the vertical shear and current at the surface employing standard HF radars.  In our settings, it yielded reliable results under wind speed exceeding 5 m/sec in average with low gustiness (<10 m/sec) and significant sea states of $H_s$>1.8m. In such circumstances, the shear value is likely to be high and measurable.  To outline more accurately the  range of parameters where this method will work reliably further studies are needed.

\noindent \textbf{Acknowledgements} This research has been supported by the Project no. 7744592 MEGIC ``Integrability and Extremal Problems in Mechanics, Geometry and Combinatorics'' of the Science Fund of Serbia and the ISRAEL SCIENCE FOUNDATION grant numbers 1601/20 and 2358/24. SS-H would like to thank for the kind support of the short-term Max-Planck Miverva post-doctoral fellowship.

\section*{Appendix: Two-Way Doppler from a Moving Reflector}
\setcounter{equation}{0}\renewcommand\theequation{A\arabic{equation}}
The two-way Doppler relation used in Eqs. (\ref{eq:dop_freq_to_vel}a)-(\ref{eq:dop_freq_to_vel}b) follows directly from the classical Doppler effect applied twice—once for transmission from the radar to the moving surface, and once for the return reflection. For a wave of frequency \(f_0\) and phase speed \(C_e\), the one-way Doppler formula is
\begin{equation}
f = \frac{C_e + v_r}{C_e - v_s}\,f_0,
\end{equation}
where \(v_r\) and \(v_s\) are the receiver and source velocities (positive when directed toward each other).

For the outgoing path, the radar acts as a stationary source (\(v_s = 0\)) and the sea-surface facet as the moving receiver (\(v_r = \pm u\)), giving
\begin{equation}
f' = \frac{C_e \pm u}{C_e}\,f_0 .
\end{equation}
On the return path, the same facet acts as a moving source (\(v_s = \pm u\)) while the radar is stationary (\(v_r = 0\)), so that
\begin{equation}
f = \frac{C_e}{C_e \mp u}\,f' .
\end{equation}
Combining both paths yields the total two-way frequency shift,
\begin{equation}
f = f_0\,\frac{C_e \pm u}{C_e \mp u}
   = f_0\,\frac{1 \pm u/C_e}{1 \mp u/C_e},
\end{equation}
where the upper sign corresponds to ocean wave motion toward (onshore) and the lower to motion away from (offshore) the radar. Solving for the reflector velocity gives
\begin{equation}
u = C_e\,\frac{f_0 - f}{f_0 + f} \quad \text{(offshore)}, 
\qquad 
u = C_e\,\frac{f - f_0}{f + f_0} \quad \text{(onshore)}.
\end{equation}
Introducing the measured peak frequencies \(f_{\text{offshore}}>0\) and \(f_{\text{onshore}}<0\) relative to \(f_0\), these expressions become
\begin{equation}
c_- = C_e\,\frac{f_0 - f_{\text{offshore}}}{f_0 + f_{\text{offshore}}}, 
\qquad
c_+ = C_e\,\frac{f_0 + f_{\text{onshore}}}{f_0 - f_{\text{onshore}}}.
\end{equation}

\section{Bibliography}
\bibliographystyle{cas-model2-names}

\bibliography{reference}

\printcredits


\end{document}